\documentclass[%
 aip,
 amsmath,amssymb,
 reprint,%
]{revtex4-2}

\usepackage{graphicx}
\usepackage{dcolumn}
\usepackage{bm}

\usepackage[utf8]{inputenc}
\usepackage[T1]{fontenc}
\usepackage{mathptmx}
\usepackage{etoolbox}
\usepackage{soul}
\usepackage{xcolor}
\sethlcolor{yellow}

\makeatletter
\def\@email#1#2{%
 \endgroup
 \patchcmd{\titleblock@produce}
  {\frontmatter@RRAPformat}
  {\frontmatter@RRAPformat{\produce@RRAP{*#1\href{mailto:#2}{#2}}}\frontmatter@RRAPformat}
  {}{}
}%
\makeatother
\begin{document}

\preprint{AIP/123-QED}

\title{Disorder-enabled directional delocalization and wave steering in time-modulated Dirac materials}
\author{Seulong Kim}
 \affiliation{
Research Institute of Basic Sciences, Ajou University, Suwon 16499, Korea
}%

\author{Kihong Kim}%
 \email{khkim@ajou.ac.kr}
\affiliation{Department of Physics, Ajou University, Suwon 16499, Korea
}%
\affiliation{School of Physics,
Korea Institute for Advanced Study, Seoul 02455, Korea
}%

\date{\today}

\begin{abstract}
We demonstrate a disorder-enabled yet localization-immune directional transport channel in time-modulated Dirac systems subject to stochastic temporal variations of a vector potential. In a spatially uniform medium, random temporal modulation induces strong Anderson localization for generic propagation directions, while waves propagating parallel to the modulation axis remain perfectly delocalized. This behavior originates from the pseudospin structure of the Dirac equation, which enforces exact suppression of interband coupling for specific propagation directions, thereby eliminating disorder-induced backscattering. As a result, temporal disorder acts as a symmetry-selective angular filter, producing highly collimated transport without spatial structuring. Unlike conventional impedance-matching–based transmission in clean time-varying media, this mechanism arises intrinsically from stochastic temporal modulation and remains robust across a wide range of disorder models. These findings establish temporal disorder as a resource for direction-selective wave control, enabling reconfigurable beam steering, adaptive filtering, and disorder-tolerant nanophotonic components such as temporal beam shapers. More broadly, this phenomenon represents a temporal analogue of disorder-induced delocalization channels known in spatially disordered systems and demonstrates that randomness, typically associated with localization and transport suppression, can instead isolate a perfectly transmitting channel through symmetry-selective dynamics.
\end{abstract}

\maketitle

\section{Introduction}

Dynamic control of wave transport in nanophotonic and electronic systems underpins key functionalities such as beam steering, routing, filtering, and adaptive signal processing. Recent advances in time-varying media have revealed a wide range of unconventional phenomena, including temporal reflection and refraction, photonic time crystals, and momentum gaps in periodically modulated systems.\cite{morg,gali,asga,nonreci,kou,akba,tv8,pac2,kouts,np1,tv2,tv6,lustig,dong,jones,ren} In parallel, stochastic temporal modulation has been shown to induce Anderson localization through disorder-driven interference, even without spatial inhomogeneity.\cite{shar,carm,garn,apf,eswa,kim2,np2} Together, these developments have significantly broadened the scope of wave manipulation beyond static material responses.

In clean time-modulated systems, exact temporal impedance-matching conditions can eliminate backward scattering at temporal interfaces, enabling reflectionless transmission for specific parameter configurations.\cite{tv9} In Dirac systems, related matching conditions arise naturally from the pseudospin structure and can be derived analytically.\cite{sk1,ok} When temporal modulation is instead stochastic, wave dynamics are typically dominated by localization effects, suppressing transport for most propagation directions. Whether disorder can instead enable robust, direction-selective transport channels in time-modulated systems remains an open question.

To place our results in context, it is important to distinguish two related but fundamentally different phenomena. The temporal Brewster effect is a deterministic mechanism: exact impedance-matching conditions at a temporal interface yield perfect transmission for specific parameter configurations, in direct analogy with spatial Brewster transmission at a planar interface.\cite{tv9} It does not rely on disorder and is absent in stochastic settings. The Brewster anomaly, by contrast, is intrinsically disorder-enabled.\cite{brewster1,jord,brewster4,kim5} Originally identified in randomly stratified spatial media, it describes the survival of a perfectly transmitting channel under strong disorder, while all other channels undergo Anderson localization. The two phenomena are therefore distinct in both origin and mechanism, and should not be conflated.

Here we demonstrate that time-modulated Dirac systems support a directional transport channel that is both disorder-enabled and immune to localization, representing a temporal analogue of disorder-induced delocalization channels associated with the Brewster anomaly. While random temporal fluctuations induce strong Anderson localization for generic propagation directions, waves propagating parallel to the modulation axis remain perfectly delocalized even under strong disorder. This behavior originates from the pseudospin structure of the Dirac equation, which enforces exact suppression of interband coupling for specific propagation directions, thereby eliminating disorder-induced backscattering. Temporal disorder thus acts as a symmetry-selective angular filter that isolates a single perfectly transmitting channel, producing highly collimated transport without any spatial structuring. The propagation direction can be dynamically controlled through the modulation axis, enabling purely temporal steering of wave transport.

Related studies have reported disorder-enabled, direction-selective suppression of localization in spatially disordered Dirac systems,\cite{fang,kk1,kk2} as well as Brewster-type total transmission in layered media with long-range randomness and in temporally periodic systems under impedance-matching conditions.\cite{lin1,lin2}
Although the present mechanism shares the disorder-enabled delocalization character of the spatial Brewster anomaly, it is rooted in pseudospin-selective channel decoupling under temporal modulation rather than polarization-dependent impedance matching at spatial interfaces. It therefore represents a distinct class of symmetry-protected, disorder-enabled transport in which randomness, typically associated with localization and suppression, acts as a filter that isolates a perfectly transmitting channel through symmetry-selective dynamics.

Beyond its fundamental significance, this mechanism enables dynamic beam steering governed entirely by temporal modulation. In electronic Dirac materials, a time-dependent vector potential can be tuned via external electric fields through $\mathbf{E}(t)=-\partial\mathbf{A}(t)/\partial t$, while in photonic Dirac lattices the same effect can be emulated through synthetic gauge-field modulation. These results establish temporal disorder as a practical design resource for reconfigurable, disorder-tolerant control of wave propagation in both photonic and electronic Dirac platforms, with direct implications for temporal beam shapers, adaptive filters, and related nanophotonic components.
This identifies a new class of disorder-enabled transport in time-modulated systems, in which localization is not universally suppressive but instead selectively eliminates all but a symmetry-protected propagation channel.

\section{Models}

We investigate wave dynamics governed by the massless pseudospin-1/2 Dirac equation in a spatially uniform medium subject to a time-dependent vector potential, $\mathbf{A}(t)=A(t)\hat{\mathbf{x}}$,
aligned along the $x$-axis with a temporally varying amplitude. The wave vector $\mathbf{k}$ forms an angle $\theta$ with the $x$-axis, which serves as the key geometric parameter characterizing the propagation direction. Spatial uniformity of the medium ensures that $\mathbf{k}$ remains constant throughout the temporal evolution, so that $\theta$ is fixed when the modulation direction is fixed; more generally, the physically relevant parameter is the angle between $\mathbf{k}$ and the instantaneous direction of $\mathbf{A}(t)$. Gauge invariance of the temporal scattering process is discussed in Supplementary Material S1.

The time-dependent Dirac equation for the two-component spinor $\Psi=(\psi_1,\psi_2)^{\mathrm{T}}$ reads
\begin{equation}
i\hbar \frac{d}{dt}\Psi(t)=
\begin{pmatrix}
0 & v_F\left(\pi_x-i\pi_y\right) \\
v_F\left(\pi_x+i\pi_y\right) & 0
\end{pmatrix}\Psi(t),
\end{equation}
with
\begin{equation}
\pi_x=\hbar k\cos\theta+eA(t), \quad \pi_y=\hbar k\sin\theta,
\end{equation}
where $k=\vert\mathbf{k}\vert$, $e$ is the electron charge, and $v_F$ is the Fermi velocity.

Eliminating $\psi_2$, the equation reduces to
\begin{equation}
\frac{d}{dt}\left[\frac{1}{\epsilon(t)}\frac{d\psi_1}{dt}\right]+\omega_0^2\epsilon^*(t)\psi_1=0,
\end{equation}
where
\begin{equation}
\epsilon(t)=e^{-i\theta}+\alpha(t), \quad \alpha(t)=\frac{eA(t)}{\hbar k}, \quad \omega_0=kv_F.
\label{eq:epsa}
\end{equation}
Here $\epsilon(t)$ is a complex parameter unrelated to electromagnetic permittivity. The corresponding group velocity is
\begin{equation}
\mathbf{v}_g=\pm v_F\left(\frac{\alpha+\cos\theta}{|\epsilon|},\frac{\sin\theta}{|\epsilon|}\right),
\end{equation}
where $\pm$ correspond to particle-like ($p$) and hole-like ($h$) branches. In general, the vector potential renders the group and phase velocities noncollinear, except for propagation along $\theta=0$ or $\pi$.

We model the vector potential as
\begin{equation}
\alpha(t)=\alpha_0+\delta\alpha(t),
\end{equation}
where $\delta\alpha(t)$ represents stochastic temporal fluctuations. To demonstrate robustness, we consider two complementary models.
In Model 1, $\delta\alpha(t)$ is Gaussian white noise with
\begin{equation}
\langle \delta\alpha(t)\delta\alpha(t')\rangle = g_0\delta(t-t'), \quad \langle \delta\alpha(t)\rangle=0,
\end{equation}
which allows analytical treatment via invariant imbedding.
In Model 2, $\delta\alpha(t)$ is piecewise constant over intervals of duration $\tau_d$, taking independent values uniformly distributed in $[-\Delta\alpha,\Delta\alpha]$. Wave propagation is obtained by averaging over many realizations.
The agreement between these models demonstrates that the disorder-enabled directional delocalization is robust and independent of the specific statistical form of temporal disorder.

\section{Methods}

\subsection{Invariant imbedding formulation}

We summarize the invariant imbedding formulation used for numerical analysis.
We consider an incident particle-like ($p$) wave of unit amplitude, $\psi_1(t)=e^{-i\omega_1 t}$ for $t\le 0$, with spatial dependence $e^{i(k_x x + k_y y)}$. When the vector potential varies within the interval $0 \le t \le T$, the wave function takes the form
\begin{equation}
\psi_1(t,T)=
\begin{cases}
e^{-i\omega_1 t}, & t < 0, \\
r(T)e^{i\omega_2 (t - T)} + s(T)e^{-i\omega_2 (t - T)}, & t > T,
\end{cases}
\end{equation}
where $r(T)$ and $s(T)$ are the temporal reflection and transmission coefficients. The frequencies $\omega_i$ ($i=1,2$) satisfy
\begin{equation}
\frac{\omega_i}{\omega_0} = |\epsilon_i| = \sqrt{1 + \alpha_i^2 + 2\alpha_i \cos\theta},
\label{eq:ss1}
\end{equation}
with $\alpha_1$ and $\alpha_2$ the initial and final normalized vector potentials.

The coefficients $r$ and $s$ describe interband ($p\to h$) and intraband ($p\to p$) transitions. Using invariant imbedding,\cite{bell,gol,ramm,klya,kk0,sk2,sk3} their evolution with respect to the modulation duration $\tau$ is governed by
\begin{equation}
\frac{1}{\omega_0}\frac{dr}{d\tau}=i\beta r+\gamma s, \quad
\frac{1}{\omega_0}\frac{ds}{d\tau}=-\gamma r-i\beta s,
\label{eq:iie}
\end{equation}
where
\begin{equation}
\beta=\frac{1+\alpha\alpha_2+(\alpha+\alpha_2)\cos\theta}{|\epsilon_2|}, \quad
\gamma=\frac{(\alpha-\alpha_2)\sin\theta}{|\epsilon_2|}.
\end{equation}
The initial conditions are
\begin{equation}
r(0)=\frac{1}{2}\left(1-\frac{\epsilon_2}{\epsilon_1}\frac{|\epsilon_1|}{|\epsilon_2|}\right), \quad
s(0)=\frac{1}{2}\left(1+\frac{\epsilon_2}{\epsilon_1}\frac{|\epsilon_1|}{|\epsilon_2|}\right),
\label{eq:ss2}
\end{equation}
and the reflectance and transmittance,
\begin{equation}
R=|r|^2, \quad S=|s|^2,
\end{equation}
satisfy the conservation law
\begin{equation}
R+S=1.
\label{eq:ss3}
\end{equation}
Detailed derivations are provided in Supplementary Material S2–S4.

Equation~(\ref{eq:iie}) has the structure of a two-level system, where $\beta$ governs intraband phase accumulation and $\gamma$ controls interband (pseudospin) coupling. Since $\gamma\propto\sin\theta$, the coupling vanishes identically at $\theta=0$ and $\pi$, leading to an exact decoupling of pseudospin channels. As a result, temporal reflection is suppressed and the dynamics remain immune to arbitrarily strong temporal disorder along these directions.

\subsection{Moments of the reflection and transmission coefficients}\label{sec3}

To characterize long-time transport under stochastic temporal modulation, we analyze disorder-averaged moments of the reflection and transmission coefficients. Instead of tracking individual realizations of $r$ and $s$, we derive closed evolution equations for their statistical moments $Z_{abcd}$. The resulting hierarchy follows directly from the linear coupling in Eq.~(\ref{eq:iie}) combined with Gaussian stochastic averaging.

Applying the Furutsu–Novikov formula to Eq.~(\ref{eq:iie}),\cite{fur,nov} we obtain a coupled system of differential equations for the moments $Z_{abcd}=\langle r^a(r^*)^{b}s^c(s^*)^{d}\rangle$:
\begin{align}
 &\frac{1}{\omega_0}\frac{d}{d\tau}Z_{abcd}=C_1Z_{abcd}+C_2Z_{a+1,b,c-1,d}\nonumber\\
   &~~~+C_3Z_{a-1,b,c+1,d}+C_4Z_{a,b+1,c,d-1}\nonumber\\
   &~~~+C_5Z_{a,b-1,c,d+1}+C_6Z_{a+1,b+1,c-1,d-1}\nonumber\\
   &~~~+C_7Z_{a-1,b-1,c+1,d+1}+C_8Z_{a+1,b-1,c-1,d+1}\nonumber\\
   &~~~+C_9Z_{a-1,b+1,c+1,d-1}+C_{10}Z_{a+2,b,c-2,d}\nonumber\\
   &~~~+C_{11}Z_{a-2,b,c+2,d}+C_{12}Z_{a,b+2,c,d-2}\nonumber\\
   &~~~+C_{13}Z_{a,b-2,c,d+2}.
   \label{eq:zmom}
\end{align}
The coefficients are
\begin{align}
&C_1=i(a-b-c+d)\beta_0-\frac{g}{2}(a-b-c+d)^2\zeta^2\nonumber\\
&~~~~~~-\frac{g}{2}(a+b+c+d+2ac+2bd)\eta^2,\nonumber\\
&C_2=-c\gamma_0-igc(a-b-c+d+1)\zeta\eta,\nonumber\\
&C_3=a\gamma_0+iga(a-b-c+d-1)\zeta\eta,\nonumber\\
&C_4=-d\gamma_0-igd(a-b-c+d-1)\zeta\eta,\nonumber\\
&C_5=b\gamma_0+igb(a-b-c+d+1)\zeta\eta,\nonumber\\
&C_6=gcd\eta^2,~~C_7=gab\eta^2,\nonumber\\
&C_8=-gbc\eta^2,~~C_9=-gad\eta^2,\nonumber\\
&C_{10}=\frac{g}{2}c(c-1)\eta^2,~~C_{11}=\frac{g}{2}a(a-1)\eta^2,\nonumber\\
&C_{12}=\frac{g}{2}d(d-1)\eta^2,~~C_{13}=\frac{g}{2}b(b-1)\eta^2,
\end{align}
where the parameters are
\begin{align}
&\beta_0=\frac{1+\alpha_0\alpha_2+(\alpha_0+\alpha_2)\cos\theta}{|\epsilon_2|},\nonumber\\
&\gamma_0=\frac{(\alpha_0-\alpha_2)\sin\theta}{|\epsilon_2|}, \quad
\zeta=\frac{\alpha_2+\cos\theta}{|\epsilon_2|},\nonumber\\
&\eta=\frac{\sin\theta}{|\epsilon_2|}, \quad g=g_0\omega_0.
\end{align}
The initial conditions are
\begin{equation}
Z_{abcd}(0) = [r(0)]^a[r^*(0)]^b[s(0)]^c[s^*(0)]^d.
\end{equation}

Two structural features are central. First, all inter-moment coupling terms are proportional to $\sin\theta$ (through $\gamma_0$ or $\eta$), so that disorder-induced mixing vanishes identically at $\theta=0$ and $\pi$. Second, the stochastic contributions scale with the disorder strength $g$, enabling systematic analysis of temporal disorder effects. As a result, the hierarchy in Eq.~(\ref{eq:zmom}) simplifies at $\theta=0$ and $\pi$ to a decoupled form with identically vanishing reflection, while for generic angles disorder induces interband mixing and moment coupling, leading to localization. In what follows, we numerically integrate the resulting finite, closed hierarchy to obtain disorder-averaged transport as functions of propagation angle and disorder strength.

\section{Results}

\begin{figure*}
  \centering
  \includegraphics[width=12cm]{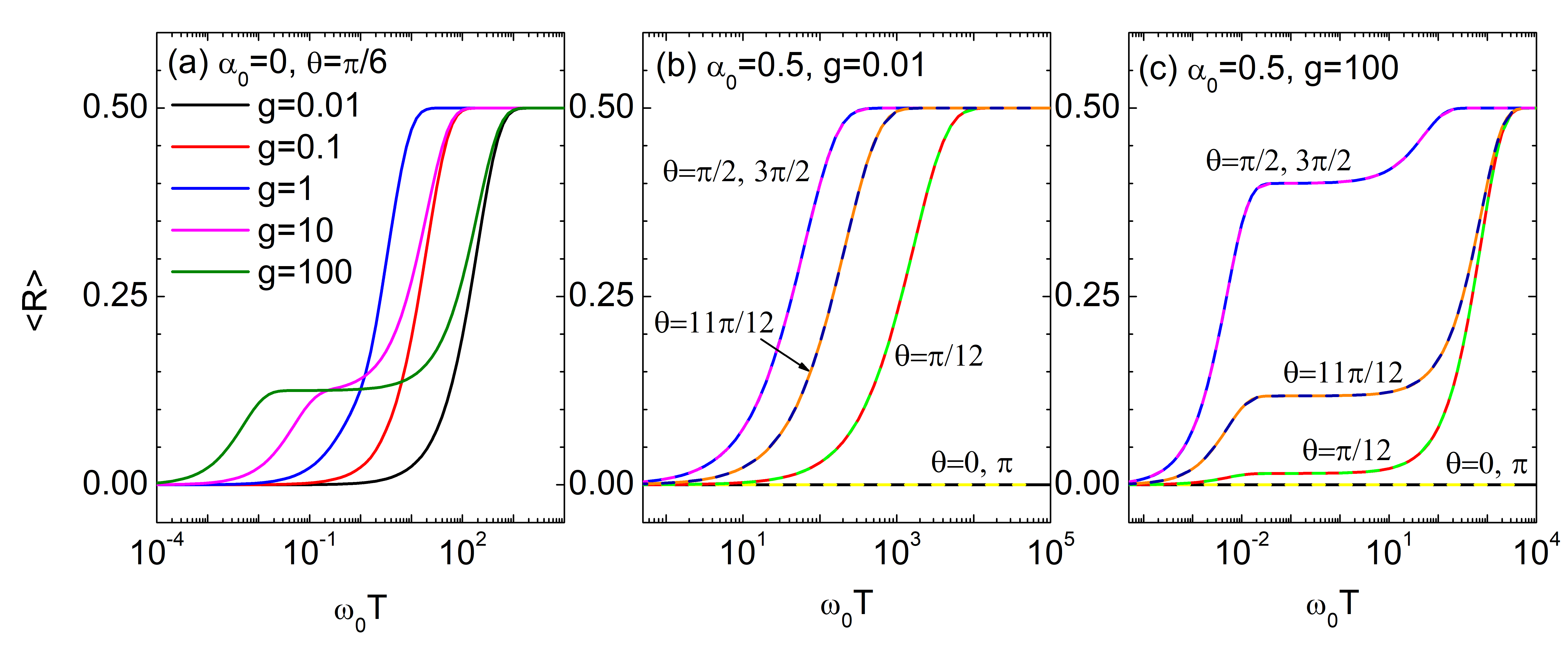}
  \caption{Time evolution of the disorder-averaged reflectance $\langle R \rangle$.
(a) $\langle R \rangle$ for several disorder strengths $g$ at fixed mean vector potential $\alpha_0 = 0$ and incidence angle $\theta = \pi/6$.
(b) $\langle R \rangle$ for different $\theta$ in the weak disorder regime ($g = 0.01$) with $\alpha_0 = 0.5$.
(c) $\langle R \rangle$ for different $\theta$ in the strong disorder regime ($g = 100$) with $\alpha_0 = 0.5$.
Solid lines represent numerical solutions of Eq.~(\ref{eq:zmom}); dashed lines show the corresponding analytical predictions from Eq.~(\ref{eq:wd}) in (b) and Eq.~(\ref{eq:sd}) in (c).
}
\label{fig1}
\end{figure*}

\begin{figure}
  \centering
  \includegraphics[width=8.3cm]{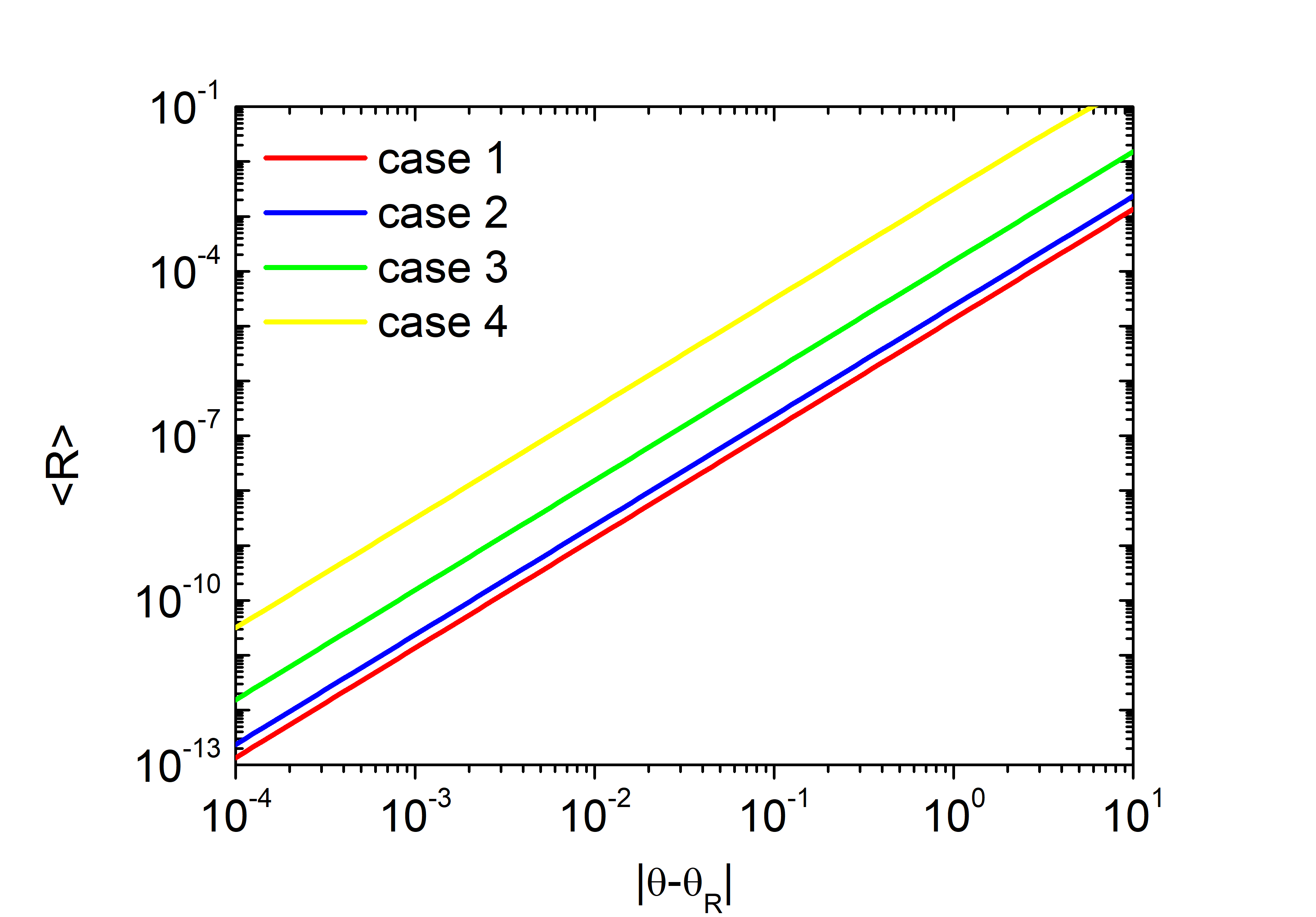}
  \caption{Quadratic scaling of the disorder-averaged reflectance near the reflectionless directions $\theta = \theta_R$ ($0$ or $\pi$).
The disorder-averaged reflectance $\langle R \rangle$ is plotted versus $\theta - \theta_R$ for four representative parameter sets:
(1) $g = 0.01$, $\alpha_0 = 0.5$, $\omega_0 T = 10$, $\theta_R = 0$;
(2) $g = 0.1$, $\alpha_0 = 0.1$, $\omega_0 T = 1$, $\theta_R = 0$;
(3) $g = 1$, $\alpha_0 = 0.2$, $\omega_0 T = 0.5$, $\theta_R = \pi$;
(4) $g = 10$, $\alpha_0 = 0$, $\omega_0 T = 100$, $\theta_R = \pi$.
In all cases, $\langle R \rangle$ exhibits the universal quadratic scaling $\langle R \rangle \propto (\theta - \theta_R)^2$ in the vicinity of $\theta_R$, demonstrating the robustness of the disorder-enabled directional delocalization} against variations in disorder strength $g$, mean vector potential $\alpha_0$, and modulation duration $T$.
\label{fig2}
\end{figure}

We analyze the temporal evolution of the ensemble-averaged reflectance $\langle R \rangle$, obtained by solving Eq.~(\ref{eq:zmom}), which is valid for arbitrary disorder strength $g$. For clarity, we focus on the symmetric case $\alpha_1=\alpha_2=\alpha_0$, where the vector potentials before and after the modulation interval coincide with their mean value during modulation.

When $\theta=0$ or $\pi$, the parameters $\gamma_0$ and $\eta$ vanish, and Eq.~(\ref{eq:zmom}) yields $\langle R\rangle=0$ and $\langle S\rangle=1$. These directions therefore correspond to disorder-immune propagation with perfect transmission independent of temporal fluctuations.
For general $\theta$, analytic expressions can be obtained in both the weak- and strong-disorder limits. In the weak-disorder regime,
\begin{eqnarray}
\left\langle R \right\rangle = \frac{1}{2}\left[1 - \exp{\left(-\frac{2 g\omega_0\sin^2{\theta}}{\left\vert\epsilon_0\right\vert^2}t\right)}\right],
\label{eq:wd}
\end{eqnarray}
where $|\epsilon_0| = \sqrt{1 + \alpha_0^2 + 2\alpha_0\cos\theta}$.
In the strong-disorder regime, numerical solutions yield
\begin{align}
\left\langle R \right\rangle =& \frac{1}{2}\left(1 - e^{-2g\omega_0t}\right)\nonumber\\&\times\left[1-\left(1-\frac{\sin^2{\theta}}{\left\vert\epsilon_0\right\vert^2}\right)
\exp{\left(-\frac{2\omega_0\sin^2{\theta}}{g}t\right)}\right].
\label{eq:sd}
\end{align}
In both limits, $\langle R \rangle$ approaches $1/2$ exponentially for any $\theta \neq 0,\pi$, indicating strong temporal localization for all propagation directions except the symmetry-protected directions $\theta=0$ and $\pi$.

Figure~\ref{fig1} shows the time evolution of $\langle R \rangle$ under representative conditions. In Fig.~\ref{fig1}(a), the incidence angle is fixed at $\theta=\pi/6$ with $\alpha_0=0$, while the disorder strength $g$ is varied. In Figs.~\ref{fig1}(b) and \ref{fig1}(c), $\alpha_0=0.5$ is fixed and $\langle R \rangle$ is shown for several angles in the weak ($g=0.01$) and strong ($g=100$) disorder regimes, respectively. In all cases, $\langle R \rangle$ increases monotonically from zero and saturates at $1/2$, except at $\theta=0$ and $\pi$, where it remains zero due to perfect transmission. In the weak-disorder regime, increasing $g$ accelerates the approach to saturation, whereas in the strong-disorder regime the asymptotic relaxation slows as $g$ increases.

Near the disorder-immune directions $\theta_R=0$ or $\pi$, the reflectance exhibits universal quadratic scaling, $\langle R \rangle \propto (\theta-\theta_R)^2$, independent of disorder strength, mean vector potential, or modulation duration. Figure~\ref{fig2} confirms this scaling across representative regimes ranging from weak to very strong disorder and from short to long modulation times.

Using the conservation law $S+R=1$, the quantity $\langle S-R\rangle=1-2\langle R\rangle$ represents the net propagation current, proportional to the current density and average group velocity.\cite{kim2} In the weak-disorder regime,
\begin{equation}
\langle S - R \rangle = e^{-t/\tau_w}, \quad
\tau_w = \frac{1 + \alpha_0^2 + 2\alpha_0\cos\theta}{2g\omega_0\sin^2\theta}.
\label{eq:wdr}
\end{equation}
In the strong-disorder regime and for large $t$,
\begin{equation}
\langle S - R \rangle \propto e^{-t/\tau_s}, \quad
\tau_s = \frac{g}{2\omega_0\sin^2\theta}.
\end{equation}
Since $g=g_0\omega_0=g_0 k v_F$, we obtain $\tau_w \propto g_0^{-1}k^{-2}$ and $\tau_s \propto g_0$, with both scaling as $(\sin\theta)^{-2}$. Thus, the net current vanishes asymptotically for all $\theta \ne 0,\pi$.

In the weak-disorder regime, stronger disorder accelerates the decay of the current. In contrast, in the strong-disorder regime the decay proceeds in two stages: an initial rapid drop to $1-\sin^2\theta/|\epsilon_0|^2$ on the timescale $(2g\omega_0)^{-1}$, followed by a slower exponential relaxation with a timescale that increases with $g$. Consequently, sufficiently strong disorder ultimately slows the asymptotic decay of the net current.

\begin{figure*}
  \centering
  \includegraphics[width=12cm]{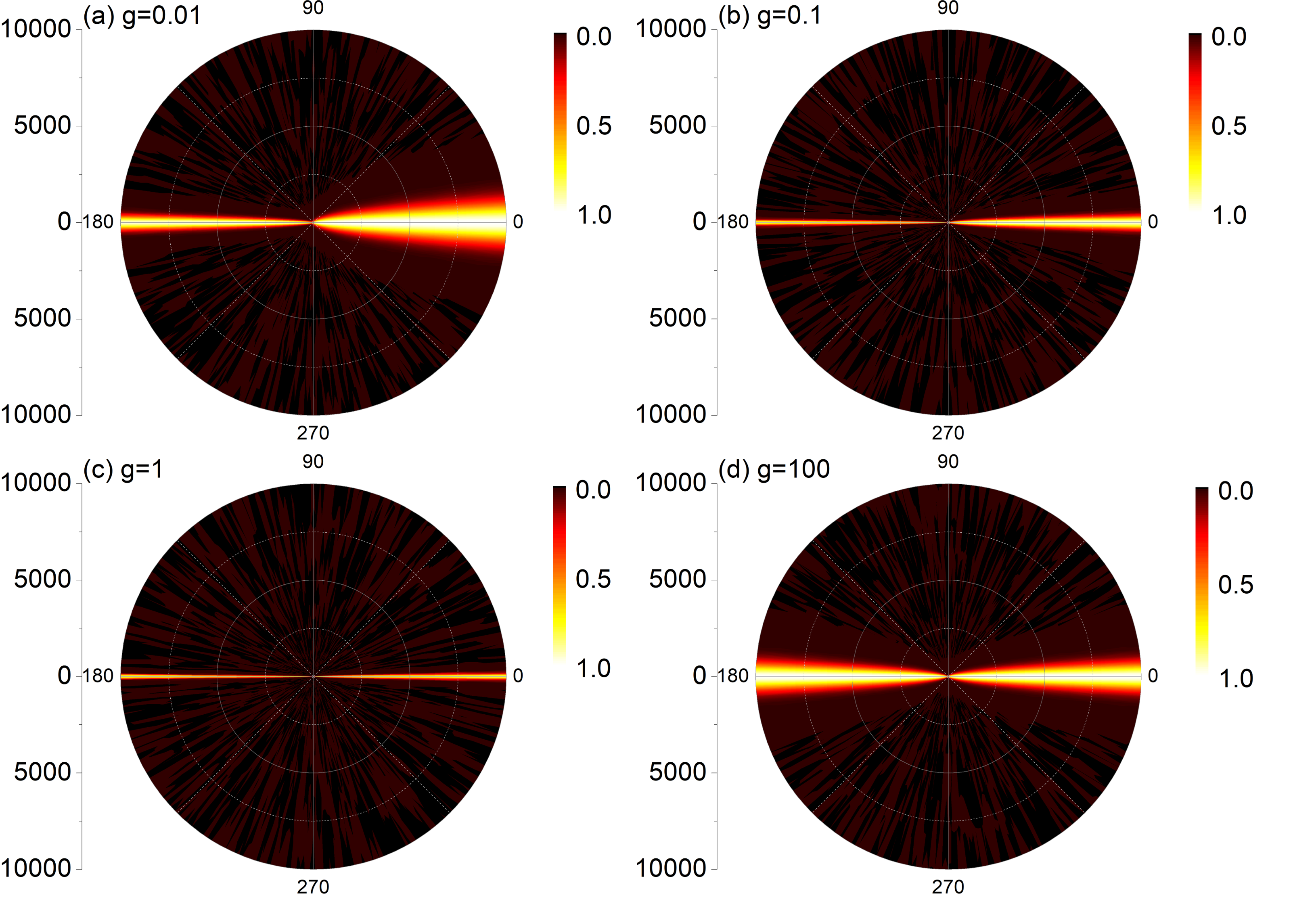}
  \caption{Polar representation of the disorder-averaged net current $\langle S - R \rangle$ under stochastic temporal modulation. The time evolution of $\langle S - R \rangle$ is shown in polar coordinates for disorder strengths (a) $g = 0.01$, (b) $0.1$, (c) $1$, and (d) $100$, spanning weak to strong disorder regimes, with fixed mean vector-potential amplitude $\alpha_0 = 0.5$. The radial coordinate represents the modulation duration (time), while the angular coordinate corresponds to the incidence angle $\theta$. The outermost radius corresponds to the maximum dimensionless duration, $T/\tau_d=10^4$. Increasing disorder progressively suppresses off-axis transport, whereas propagation near the reflectionless directions remains robust, demonstrating robust, disorder-immune directional transport.}
    \label{fig3}
\end{figure*}

Figure~\ref{fig3} presents the temporal evolution of the net current $\langle S - R \rangle$ in polar coordinates for various disorder strengths $g$, with the mean vector potential fixed at $\alpha_0=0.5$. The radial coordinate represents time, while the angular coordinate corresponds to the incidence angle $\theta$ between the wave vector and the vector-potential axis. Finite values of $\langle S - R \rangle$ indicate sustained propagation, whereas values approaching zero signal localization and suppression of transport.

As time increases, $\langle S - R \rangle$ decays from 1 to 0 for nearly all propagation directions, except within narrow angular regions near $\theta=0$ and $\pi$, where transmission persists. These symmetry-protected directions remain disorder-immune and define the directional transport channel identified in this work. In the weak-disorder regime, the angular window supporting propagation narrows as $g$ increases, whereas in the strong-disorder regime it broadens with further increase of $g$. In addition, under weak disorder the decay rate depends asymmetrically on $\alpha_0$ and $\theta$ [see Eq.~(\ref{eq:wdr})], leading to directional asymmetry between $\theta=0$ and $\pi$, which disappears in the strong-disorder limit.

\begin{figure}
  \centering
  \includegraphics[width=8.3cm]{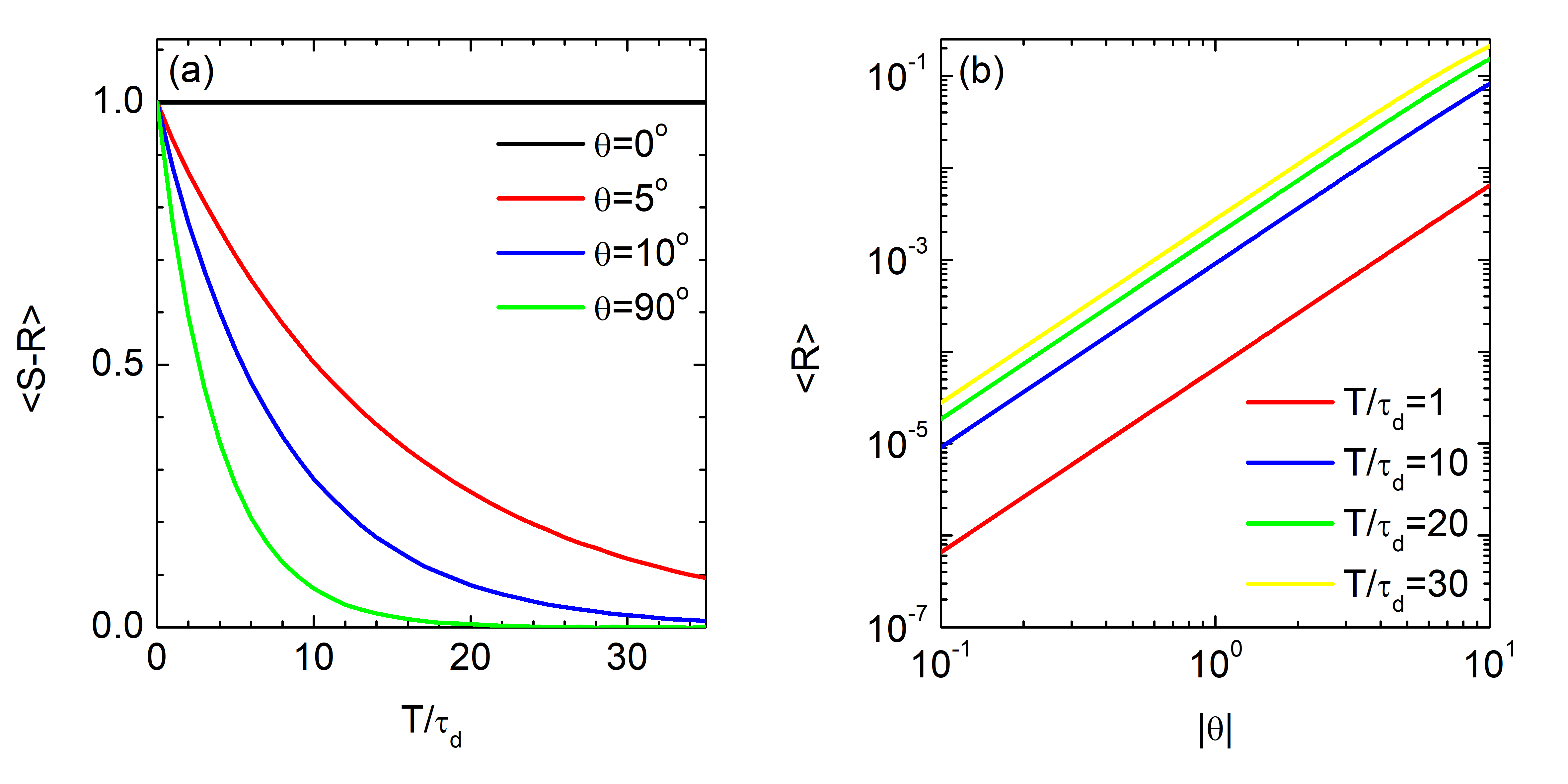}
  \caption{Directional wave localization and reflectance in Model 2.
(a) Time evolution of the disorder-averaged net current $\langle S - R \rangle$ for several incidence angles $\theta$, obtained from $10^6$ stochastic realizations with dimensionless modulation amplitude $\Delta\alpha = 1$, mean vector-potential amplitude $\alpha_0 = 0$, and dimensionless correlation time $\omega_0\tau_d = 10$, representative of typical modulation conditions. The current remains unity at $\theta = 0$, evidencing a disorder-immune propagation channel, whereas for $\theta \neq 0$ it decays exponentially in time, indicating Anderson localization induced by temporal disorder.
(b) Disorder-averaged reflectance $\langle R \rangle$ versus $|\theta|$ for different modulation durations. In the small-angle regime, $\langle R \rangle \propto \theta^2$, demonstrating quadratic suppression near $\theta = 0$ and confirming robust directional selectivity.
}
\label{fig4}
\end{figure}

We next examine wave propagation in Model 2. Despite the different statistical structure of the disorder, the qualitative behavior remains unchanged. Figure~\ref{fig4}(a) shows the time evolution of $\langle S - R \rangle$ for several incidence angles $\theta$, computed with $\Delta\alpha=1$, $\alpha_0=0$, and $\omega_0\tau_d=10$, and averaged over $10^6$ realizations. For $\theta=0$, temporal scattering is absent and the net current remains unity, while for $\theta\neq 0$ it decays exponentially, indicating suppression of forward transport due to Anderson localization induced by temporal disorder.
Figure~\ref{fig4}(b) shows the disorder-averaged reflectance $\langle R \rangle$ as a function of $|\theta|$ for several modulation durations. As in Model 1, $\langle R \rangle$ exhibits quadratic scaling near $\theta=0$, $\langle R \rangle \propto \theta^2$, confirming the persistence of the disorder-immune propagation direction.

These results demonstrate that disorder-induced temporal modulation produces directional localization for all $\theta\neq 0,\pi$, independent of the disorder model, while delocalized transport persists only along the symmetry-protected directions $\theta=0$ and $\pi$. This behavior is further illustrated by direct wave-packet simulations.

To this end, we consider a two-dimensional pulse initially centered at the origin with a Gaussian momentum distribution,
\begin{align}
u(x,y,t)=\int_{-\infty}^{\infty}\int_{-\infty}^{\infty} D(k_x,k_y)e^{ik_xx+ik_yy}\Psi(k_x,k_y,t), dk_xdk_y,
\label{eq:pul1}
\end{align}
where, for $t>T$,
\begin{align}
&\Psi(k_x,k_y,t)=s(k_x,k_y,T)\begin{pmatrix} 1\\ \epsilon_2^*(k_x,k_y)/\left\vert\epsilon_2(k_x,k_y)\right\vert\end{pmatrix} e^{-i\omega_2\left(t-T\right)}\nonumber\\
&~~~~~~~~~~~~~~~~+r(k_x,k_y,T)\begin{pmatrix} 1\\ -\epsilon_2^*(k_x,k_y)/\left\vert\epsilon_2(k_x,k_y)\right\vert\end{pmatrix} e^{i\omega_2\left(t-T\right)},\nonumber\\
&D(k_x,k_y)=e^{-\frac{k_x^2+k_y^2}{2\sigma_k^2}},
\quad \omega_2=\vert \epsilon_2\vert kv_F,
\label{eq:pul2}
\end{align}
and $\epsilon_2$ is evaluated at the final time corresponding to the modulation duration $T$. The normalized field intensity is defined as
\begin{equation}
F(x,y,T)=\frac{|u(x,y,T)|^2}{\int |u(x,y,0)|^2, dxdy}.
\label{eq:pul3}
\end{equation}

\begin{figure*}
  \centering\includegraphics[width=6cm]{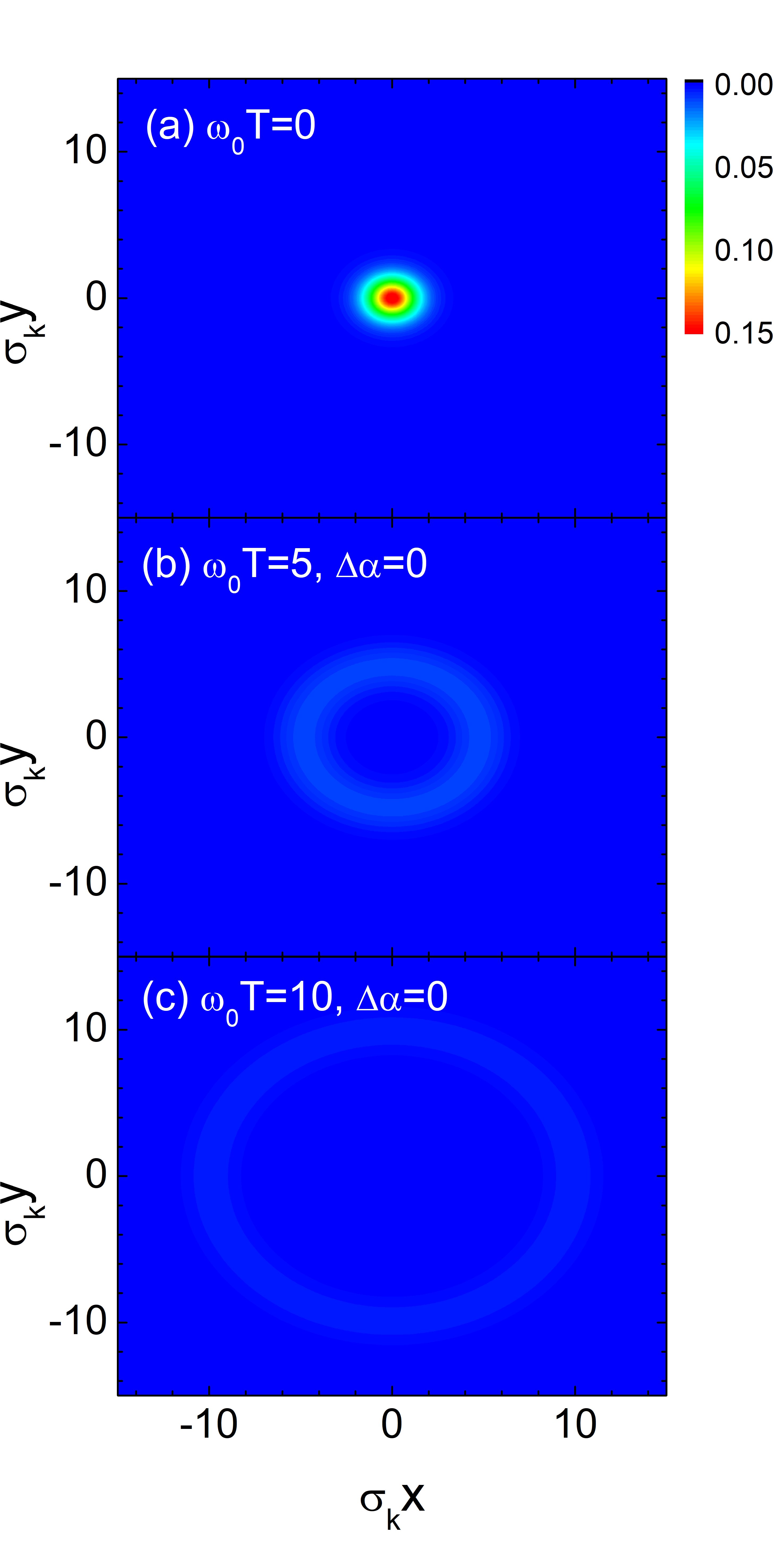}\includegraphics[width=6cm]{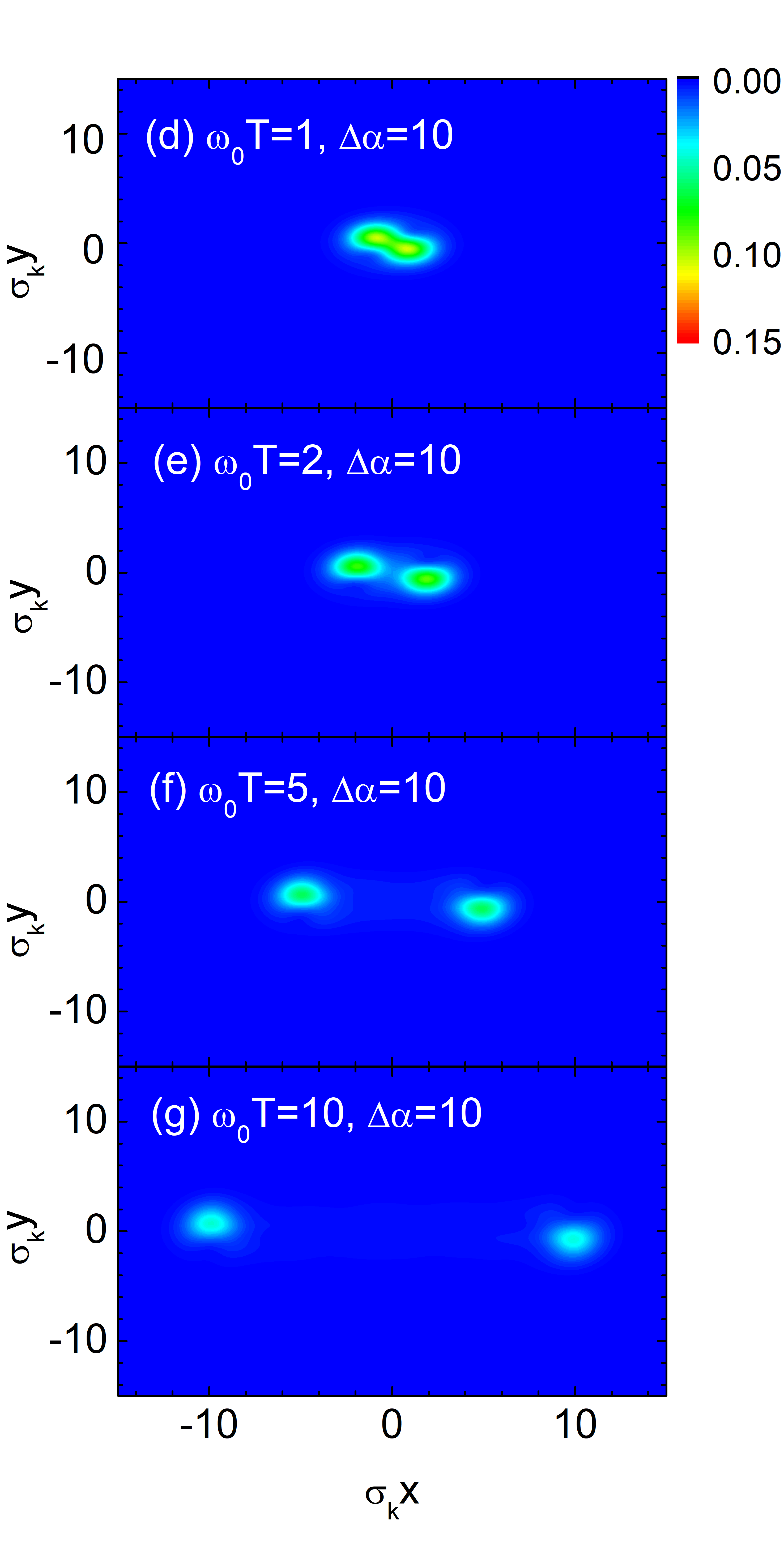}
  \caption{Temporal evolution of a two-dimensional pulse in stationary and time-disordered media.
(a–c) In a stationary medium, the pulse expands isotropically, spreading uniformly in all directions.
(d–g) Under stochastic temporal modulation of the vector potential along the $x$-axis with dimensionless modulation amplitude $\Delta\alpha = 10$, corresponding to a strong-modulation regime, the Dirac wave becomes directionally localized and propagates predominantly along the $\pm x$ directions, leading to the formation of two symmetric lobes.
All results are averaged over 200 disorder realizations with correlation-time–bandwidth product $v_F\sigma_k\tau_d = 1$, for which the disorder memory time is comparable to the inverse spectral bandwidth of the pulse.}
    \label{fig5}
\end{figure*}

\begin{figure*}
  \centering\includegraphics[width=12cm]{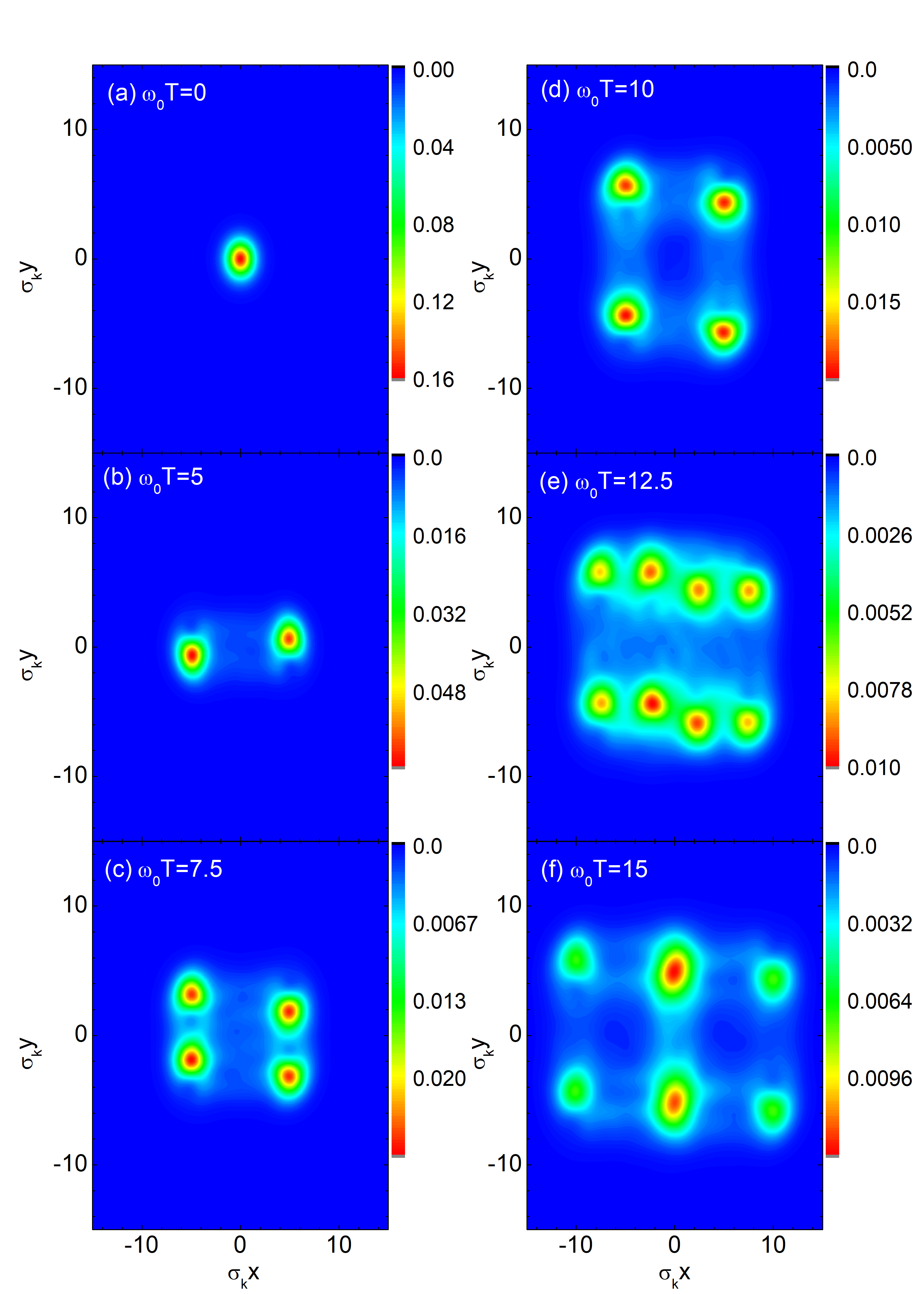}
  \caption{Steering of a Dirac pulse via directional modulation of a time-disordered vector potential.
A Dirac pulse propagates in a medium where the orientation of the temporally disordered vector potential alternates between the $x$- and $y$-axes at intervals of $\omega_0 T = 5$, chosen to allow directional lobe formation before switching. The disorder strength is $\Delta\alpha = 10$, corresponding to a strong-modulation regime that enables clear directional splitting. All results are averaged over 200 random realizations with correlation-time–bandwidth product $v_F\sigma_k\tau_d = 1$.
(a) The initial Gaussian pulse is centered at the origin.
(b) For $0 < \omega_0 T < 5$, the vector potential is aligned along $x$, steering the pulse along $\pm x$ and forming two lobes.
(c,d) For $5 < \omega_0 T < 10$, the direction switches to $y$, redirecting the pulse along $\pm y$ and producing four lobes.
(e) For $10 < \omega_0 T < 15$, the orientation returns to $x$, further splitting the pulse into eight lobes.
(f) At $\omega_0 T = 15$, interference near $x = 0$ leads to partial merging, resulting in six distinct lobes.}
    \label{fig6}
\end{figure*}

Figure~\ref{fig5} illustrates the temporal evolution of a two-dimensional pulse initially centered at the origin, composed of wave vectors with an isotropic Gaussian distribution. In a stationary medium, the pulse expands uniformly in all directions, forming a circular front whose amplitude decreases due to spatial spreading.

When the vector potential undergoes stochastic temporal modulation with strength $\Delta\alpha=10$ and fixed orientation along the $x$-axis, the Dirac wave becomes directionally localized and propagates only along $\theta=0$ and $\pi$. The pulse therefore splits into two components traveling along the $\pm x$ directions. This strong directional confinement provides a direct real-space manifestation of the disorder-immune transport channel identified above. Although the overall amplitude decreases with time, confinement to two propagation channels substantially slows the decay, allowing the pulse to retain appreciable intensity over extended durations. The resulting two sharply defined propagation lobes demonstrate disorder-selected transport channels, in which temporal fluctuations actively shape the wavefront rather than degrade it.

Figure~\ref{fig6} further demonstrates that the propagation direction can be dynamically controlled in time. In this example, the orientation of the temporally disordered vector potential alternates between the $x$- and $y$-axes every $\omega_0 T=5$, with $\Delta\alpha=10$, and results averaged over 200 disorder realizations for $\sigma_k \tau_d=1$. During the first interval ($0<\omega_0 T<5$), the pulse splits into two lobes along $\pm x$. In the next interval ($5<\omega_0 T<10$), the propagation direction switches to $\pm y$, producing four lobes. When the orientation returns to $x$ ($10<\omega_0 T<15$), the pulse further divides into eight lobes. At $\omega_0 T=15$, interference near $x=0$ leads to partial merging, resulting in six distinct lobes.

These results show that temporal control of the vector-potential orientation enables dynamic beam steering and real-time reconfiguration of transport pathways, providing a direct real-space manifestation of disorder-immune directional transport. More broadly, temporal modulation enables programmable routing of wave energy through disorder itself, allowing reconfiguration of propagation pathways without any modification of the underlying spatial structure.

\section{Discussion}

In contrast to conventional wave transport, where disorder typically degrades coherence and suppresses transmission, the present system demonstrates that temporal randomness can instead enforce directional selectivity by dynamically suppressing competing scattering pathways. Temporal disorder thus acts as a resource that isolates a symmetry-protected transport channel.

This mechanism can be realized in electronic Dirac materials such as graphene, where a time-dependent vector potential is generated by an in-plane electric field via $\mathbf{E}(t)=-\partial\mathbf{A}(t)/\partial t$.\cite{mendon} For representative parameters, taking $v_F \sim 10^{6}$ m/s and a carrier energy of $\sim 0.1$ eV yields a wavenumber $k \approx 1.5\times 10^8$ m$^{-1}$.\cite{neto,peres,ploc,lher} The linear Dirac dispersion extends over several hundred meV,\cite{ploc} ensuring the validity of this estimate within the massless regime. For $\Delta\alpha = 1$ (with $\alpha = eA/(\hbar k)$), the corresponding vector-potential variation is $\Delta A \approx 10^{-7}$ V·s/m. Imposing $\omega_0 \tau_D \sim 10$ then requires an electric field amplitude on the order of $10^{-4}$ V/Å, which lies within demonstrated ultrafast optical capabilities. These parameters remain tunable via carrier energy and modulation bandwidth; at lower carrier energies, the characteristic timescale increases, enabling picosecond-scale gating while maintaining the condition $\omega_0\tau_D \approx 10$.

A complementary realization is provided by photonic Dirac systems, such as honeycomb coupled-resonator (photonic graphene) lattices supporting Dirac cones.\cite{rech,plot} In these platforms, synthetic gauge fields can be implemented through dynamic modulation of hopping phases.\cite{fan} Electro-optic phase control in thin-film lithium-niobate and related integrated platforms enables in-situ, high-bandwidth modulation of the coupling phase.\cite{cwang} Using $\alpha=eA/(\hbar k)$, the effective vector potential maps to a Peierls phase $\Delta\phi$, yielding $\Delta\alpha \sim \Delta\phi/(qa)$ near the Dirac point. For typical values $qa \sim 0.1$–1, the regime $\Delta\alpha \gtrsim 1$ corresponds to $\Delta\phi \sim 0.1$–1 rad, well within current electro-optic phase modulation capabilities. Temporal disorder can be implemented via programmable phase sequences, with required modulation speeds compatible with sub-nanosecond electronics.

The analysis further shows that the disorder-immune transport channel is fully robust in the massless case, persisting even under random fluctuations of the Fermi velocity. In the presence of a finite mass, the effect is recovered when variations of the vector potential, Fermi velocity, and mass satisfy an appropriate matching condition; otherwise, small mass fluctuations introduce only perturbative corrections (see Supplementary Material S5).

The underlying mechanism is intrinsically tied to the pseudospin structure of Dirac systems. Interband coupling vanishes identically along specific propagation directions, leading to exact decoupling of transport channels and complete suppression of backscattering. This symmetry-protected channel persists even under strong temporal disorder and remains robust against parameter fluctuations.
By contrast, in conventional electromagnetic systems, random temporal modulation typically induces parametric growth of wave energy at all angles, obscuring any disorder-selective transparent channel.\cite{shar,carm,np2} This highlights a fundamental distinction between Dirac and non-Dirac wave systems.

These results establish temporal disorder as a controllable design parameter for direction-selective transport. By exploiting symmetry-selective dynamics, one can realize functionalities such as dynamic beam steering, adaptive filtering, and reconfigurable routing without relying on spatial structuring. More broadly, this work demonstrates that randomness, when coupled to internal symmetries, can enhance rather than degrade functionality, providing a new paradigm for wave control in time-modulated systems.

\begin{acknowledgments}
This research was supported by the National Research Foundation of Korea (NRF) grant funded by the Korean Government (RS-2025-16071339). It was also supported by the Basic Science Research Program through the NRF, funded by the Ministry of Education (RS-2021-NR060141).
\end{acknowledgments}

\section*{Data Availability}

The data that support the findings of this study are available
from the corresponding author upon reasonable request.

\newpage
\setcounter{equation}{0}
\renewcommand{\theequation}{S\arabic{equation}}
\setcounter{figure}{0}
\renewcommand{\thefigure}{S\arabic{figure}}

\bigskip
\bigskip

\begin{widetext}
\section*{S1. Gauge invariance of temporal scattering}

The pseudospin-1/2 Dirac equation with electromagnetic potentials reads
\begin{equation}
 i\hbar\frac{\partial}{\partial t}\Psi=v_F\left[\left(\frac{\hbar}{i} \frac{\partial}{\partial x}+eA_x\right)\sigma_x+\left(\frac{\hbar}{i} \frac{\partial}{\partial y}+eA_y\right)\sigma_y\right]\Psi-eV\Psi.
 \end{equation}
Under a gauge transformation,
 \begin{equation}
 A\to A+\nabla\Lambda,~V\to V-\frac{\partial\Lambda}{\partial t},
 \end{equation}
and with the transformation of the wave function,
\begin{equation}
\Psi\to\Psi^\prime=e^{i\lambda}\Psi,
\end{equation}
the additional terms cancel provided
\begin{equation}
\lambda=-\tfrac{e}{\hbar}\Lambda,
\end{equation}
thus ensuring gauge invariance.

Defining the effective wave vector ${\bf q}={\bf k}+e{\bf A}/\hbar$ through the relation
\begin{equation}
\left(\tfrac{\hbar}{i}\tfrac{\partial}{\partial x}+eA_x\right)\Psi
=(\hbar k_x+eA_x)\Psi=\hbar q_x\Psi,
\end{equation}
one finds that $q_x$ (and, analogously, $q_y$) is invariant under the gauge transformation.
Since the temporal scattering dynamics depend only on the gauge-invariant combination $\epsilon \propto q_x - i q_y$, both the temporal scattering process and the associated directional decoupling condition are gauge-invariant.

\section*{S2. Derivation of Eq.~(9)}

For the time-independent case, assuming a time dependence of $\exp(-i\omega t)$, Eq.~(1) can be written as
\begin{equation}
i\hbar\frac{d}{dt}\Psi(t)=\hbar\omega\Psi(t)=\hbar v_F\begin{pmatrix}
                     0 &  k \exp^{-i\theta}+\frac{eA}{\hbar} \\
                     k \exp^{i\theta}+\frac{eA}{\hbar} & 0
                   \end{pmatrix}\Psi(t).
                   \label{eq:s1}
\end{equation}
This may be rearranged into
\begin{equation}
\begin{pmatrix}
                     \omega &  -kv_F\left(\exp^{-i\theta}+\alpha\right) \\
                     -kv_F\left(\exp^{i\theta}+\alpha\right) & \omega
                   \end{pmatrix}\Psi\equiv M\Psi=0,
                   \label{eq:s2}
\end{equation}
where
\begin{equation}
\alpha=\frac{eA}{\hbar k}.
\end{equation}
A nontrivial solution requires $\det(M)=0$. Evaluating the determinant yields Eq.~(9):
\begin{equation}
\omega=\pm kv_F\sqrt{1+\alpha^2+2\alpha\cos{\theta}}.
\end{equation}
Introducing the complex parameter
\begin{equation}
\epsilon=e^{-i\theta}+\alpha,
\end{equation}
we obtain
\begin{align}
    \frac{\omega}{\omega_0}=\vert\epsilon\vert,~~\omega_0\equiv kv_F.
\end{align}

\section*{S3. Derivation of Eq.~(12)}

Consider a temporal interface at $t=0$, where $\epsilon$ changes abruptly from $\epsilon_1$ to $\epsilon_2$.
From Eq.~(1), the spinor $\Psi(t)$ must remain continuous across the interface.
Equation~(\ref{eq:s1}) then gives the relation between $\psi_1$ and $\psi_2$:
\begin{equation}
\psi_2=\frac{i}{kv_F\epsilon}\frac{d\psi_1}{dt}.
\label{eq:s7}
\end{equation}
For an interface of zero duration ($T=0$), combination with Eq.~(8) gives the boundary conditions
\begin{align}
 &s(0)+r(0)=1,\nonumber\\
 &\frac{\omega_2}{\epsilon_2}\left[s(0)-r(0)\right]=
\frac{\omega_1}{\epsilon_1}.
\end{align}
With $\omega_i=\vert\epsilon_i\vert\omega_0$ ($i=1,2$), we arrive at Eq.~(12):
\begin{equation}
r(0)=\frac{1}{2}\left(1-\frac{\epsilon_2}{\epsilon_1}\frac{\vert\epsilon_1\vert}{\vert\epsilon_2\vert}\right),
~~s(0)=\frac{1}{2}\left(1+\frac{\epsilon_2}{\epsilon_1}\frac{\vert\epsilon_1\vert}{\vert\epsilon_2\vert}\right).
\end{equation}

\section*{S4. Proof of the conservation law $S+R=1$}

Reflectance and transmittance are defined as
\begin{equation}
R=|r|^2,~~S=|s|^2.
\end{equation}
From Eq.~(10), it follows that
\begin{align}
\frac{d}{d\tau}(S+R)
=s^*\frac{ds}{d\tau}+s\frac{ds^*}{d\tau}+r^*\frac{dr}{d\tau}+r\frac{dr^*}{d\tau}=0,
\end{align}
since $\beta$ and $\gamma$ are real. Therefore, the sum $S+R$ is conserved.
At a single interface ($T=0$), we have
\begin{align}
S(0)+R(0)&=|s(0)|^2+|r(0)|^2\nonumber\\
&=\frac{1}{4}\left(1+\frac{\epsilon_2}{\epsilon_1}\frac{\vert\epsilon_1\vert}{\vert\epsilon_2\vert}\right)
\left(1+\frac{\epsilon_2^*}{\epsilon_1^*}\frac{\vert\epsilon_1\vert}{\vert\epsilon_2\vert}\right)
+\frac{1}{4}\left(1-\frac{\epsilon_2}{\epsilon_1}\frac{\vert\epsilon_1\vert}{\vert\epsilon_2\vert}\right)
\left(1-\frac{\epsilon_2^*}{\epsilon_1^*}\frac{\vert\epsilon_1\vert}{\vert\epsilon_2\vert}\right)\nonumber\\
&=\frac{1}{4}\left(2+\frac{\epsilon_2}{\epsilon_1}\frac{\vert\epsilon_1\vert}{\vert\epsilon_2\vert}
+\frac{\epsilon_2^*}{\epsilon_1^*}\frac{\vert\epsilon_1\vert}{\vert\epsilon_2\vert}\right)
+\frac{1}{4}\left(2-\frac{\epsilon_2}{\epsilon_1}\frac{\vert\epsilon_1\vert}{\vert\epsilon_2\vert}
-\frac{\epsilon_2^*}{\epsilon_1^*}\frac{\vert\epsilon_1\vert}{\vert\epsilon_2\vert}\right)\nonumber\\
&=1.
\end{align}
Hence, $S+R=1$ always holds.

\section*{S5. Effect of Fermi-velocity and mass fluctuations on disorder-enabled directional delocalization}

We recall that temporal scattering across an interface (from medium 1 to 2) in pseudospin-1/2 Dirac systems is governed by the parameter
\begin{equation}
f=\frac{\mu_1\mu_2+v_{x1}v_{x2}q_{x1}q_{x2}+v_{y1}v_{y2}q_{y1}q_{y2}}{\sqrt{\mu_1^2+v_{x1}^2q_{x1}^2+v_{y1}^2q_{y1}^2}\sqrt{\mu_2^2+v_{x2}^2q_{x2}^2+v_{y2}^2q_{y2}^2}},
\end{equation}
where $v_{xi}$ and $v_{yi}$ are the anisotropic Fermi-velocity components along $x$ and $y$, $\mu_i=M_i/\hbar$ with $M_i$ the mass energy, and ${\bf q}={\bf k}+e{\bf A}/\hbar$ [Eq.~(20) of Ref.~25]. Directional delocalization arises when $f=\pm 1$, even in the presence of temporal disorder. In this case, scattering corresponds to a total intraband or total interband transition, with no mixing between channels, thereby guaranteeing robust delocalized transport.

For definiteness, let the vector potential be oriented along $x$ ($A_y=0$). In this case, the directional delocalization condition can be realized only when $k_y=0$.
In the massless case ($\mu_1=\mu_2=0$), one obtains
\begin{equation}
f=\frac{v_{x1}v_{x2}q_{x1}q_{x2}} {\sqrt{v_{x1}^2q_{x1}^2}\sqrt{v_{x2}^2q_{x2}^2}}
={\rm sgn}(v_{x1}q_{x1}){\rm sgn}(v_{x2}q_{x2})=\pm1,
\end{equation}
showing that the directional delocalization persists exactly, even in the presence of temporal randomness of the Fermi velocity $v_x$.

For finite $\mu$ and $k_y=0$, the scattering parameter becomes
\begin{equation}
f=\frac{\mu_1\mu_2+v_{x1}v_{x2}q_{x1}q_{x2}}
{\sqrt{\mu_1^2+v_{x1}^2q_{x1}^2}\sqrt{\mu_2^2+v_{x2}^2q_{x2}^2}},
\end{equation}
which equals $\pm1$ only if
\begin{equation}
\frac{v_{x1}q_{x1}}{\mu_1}=\frac{v_{x2}q_{x2}}{\mu_2}.
\label{eq:www}
\end{equation}
Thus, in massive systems the directional delocalization is not generally realized under arbitrary disorder, but it can still occur when the temporal variations of the vector potential, Fermi velocity, and mass are correlated so as to satisfy Eq.~(\ref{eq:www}).

If Eq.~(\ref{eq:www}) is not satisfied, the exact directional delocalization is lost because the additional $\mu$-dependent terms prevent $f=\pm 1$ in general. This is analogous to the massless case at finite $k_y$, where perfect cancellation is also absent. Nevertheless, for small $\mu$ the deviation from $\vert f\vert=1$ is only quadratic:
\begin{equation}
f\approx {\rm sgn}(v_{x1}q_{x1}){\rm sgn}(v_{x2}q_{x2})\left(1+\frac{\mu_1\mu_2}{v_{x1}v_{x2}q_{x1}q_{x2}}-\frac{\mu_1^2}{2v_{x1}^2q_{x1}^2}-\frac{\mu_2^2}{2v_{x2}^2q_{x2}^2}\right).
\end{equation}
Along the directional delocalization direction $(k_y=0)$, this correction is minimized, so scattering remains weakest in that direction.

In summary, although a mass gap generally removes the exact condition for directional delocalization, the suppression of backscattering remains nearly intact for moderate masses. As a result, transport retains the essential features of the massless Dirac limit, with only perturbative deviations.

\end{widetext}

\begin{thebibliography}{9}

\bibitem{morg}
F.~R.~Morgenthaler, ``Velocity modulation of electromagnetic waves,''
IRE Trans. Microwave Theory Tech. \textbf{6}, 167--172 (1958).

\bibitem{gali}
E. Galiffi, R. Tirole, S. Yin, H. Li, S. Vezzoli, P. A. Huidobro,
M. G. Silveirinha, R. Sapienza, A. Al\`u, and J. B. Pendry, ``Photonics of time-varying media,''
Adv. Photonics \textbf{4}, 014002 (2022).

\bibitem{asga}
M.~M.~Asgari, P.~Garg, X.~Wang, M.~S.~Mirmoosa, C.~Rockstuhl, and V.~Asadchy,
``Theory and applications of photonic time crystals: a tutorial,''
Adv. Opt. Photonics \textbf{16}, 958--1063 (2024).

\bibitem{nonreci}
D.~L.~Sounas and A.~Al\`u, ``Non-reciprocal photonics based on time modulation,''
Nat. Photonics \textbf{11}, 774--783 (2017).

\bibitem{kou}
T.~T.~Koutserimpas and R.~Fleury, ``Electromagnetic waves in a time periodic medium with step-varying refractive index,''
IEEE Trans. Antennas Propag. \textbf{66}, 5300--5307 (2018).

\bibitem{akba}
A.~Akbarzadeh, N.~Chamanara, and C.~Caloz, ``Inverse prism based on temporal discontinuity and spatial dispersion,''
Opt. Lett. \textbf{43}, 3297--3300 (2018).

\bibitem{tv8}
V.~Pacheco-Pe\~na and N.~Engheta, ``Temporal aiming,''
Light Sci. Appl. \textbf{9}, 129 (2020).

\bibitem{pac2}
V.~Pacheco-Pe\~na and N.~Engheta, ``Antireflection temporal coatings,''
Optica \textbf{7}, 323--331 (2020).

\bibitem{kouts}
T.~T.~Koutserimpas and F.~Monticone, ``Time-varying media, dispersion, and the principle of causality,''
Opt. Mater. Express \textbf{14}, 1222--1236 (2024).

\bibitem{np1}
S.~Kim and K.~Kim, ``Deterministic time rewinding of waves in time-varying media,''
Nanophotonics \textbf{14}, 3287--3298 (2025).

\bibitem{tv2}
Y. Zhou, M. Z. Alam, M. Karimi, J. Upham, O. Reshef, C. Liu,
A. E. Willner, and R. W. Boyd, ``Broadband frequency translation through time refraction in an epsilon-near-zero material,''
Nat. Commun. \textbf{11}, 2180 (2020).

\bibitem{tv6}
H.~Moussa, G.~Xu, S.~Yin, E.~Galiffi, Y.~Ra’di, and A.~Al\`u,
``Observation of temporal reflection and broadband frequency translation at photonic time interfaces,''
Nat. Phys. \textbf{19}, 863--868 (2023).

\bibitem{lustig}
E. Lustig, O. Segal, S. Saha, E. Bordo, S. N. Chowdhury, Y. Sharabi,
A. Fleischer, A. Boltasseva, O. Cohen, V. M. Shalaev, and M. Segev, ``Time-refraction optics with single cycle modulation,''
Nanophotonics \textbf{12}, 2221--2230 (2023).

\bibitem{dong}
Z.~Dong, H.~Li, T.~Wan, Q.~Liang, Z.~Yang, and B.~Yan,
``Quantum time reflection and refraction of ultracold atoms,''
Nat. Photonics \textbf{18}, 68--73 (2024).

\bibitem{jones}
T.~R.~Jones, A.~V.~Kildishev, M.~Segev, and D.~Peroulis,
``Time-reflection of microwaves by a fast optically-controlled time-boundary,''
Nat. Commun. \textbf{15}, 6786 (2024).

\bibitem{ren}
Y. Ren, K. Ye, Q. Chen, F. Chen, L. Zhang, Y. Pan,
W. Li, X. Li, L. Zhang, H. Chen, and Y. Yang, ``Observation of momentum-gap topology of light at temporal interfaces in a time-synthetic lattice,''
Nat. Commun. \textbf{16}, 707 (2025).

%

\bibitem{shar}
Y.~Sharabi, E.~Lustig, and M.~Segev, ``Disordered photonic time crystals,''
Phys. Rev. Lett. \textbf{126}, 163902 (2021).

\bibitem{carm}
R.~Carminati, H.~Chen, R.~Pierrat, and B.~Shapiro,
``Universal statistics of waves in a random time-varying medium,''
Phys. Rev. Lett. \textbf{127}, 094101 (2021).

\bibitem{garn}
J.~Garnier, ``Wave propagation in periodic and random time-dependent media,''
Multiscale Model. Simul. \textbf{19}, 1190--1211 (2021).

\bibitem{apf}
B.~Apffel, S.~Wildeman, A.~Eddi, and E.~Fort,
"Experimental implementation of wave propagation in disordered time-varying media,"
Phys. Rev. Lett. \textbf{128}, 094503 (2022).

\bibitem{eswa}
K.~S.~Eswaran, A.~E.~Kopaei, and K.~Sacha,
``Anderson localization in photonic time crystals,''
Phys. Rev. B \textbf{111}, L180201 (2025).

\bibitem{kim2}
S.~Kim and K.~Kim,
``Spatial localization and diffusion of Dirac particles and waves induced by random temporal medium variations,''
Commun. Phys. \textbf{8}, 32 (2025).

\bibitem{np2}
S.~Kim and K.~Kim, ``Statistical regimes of electromagnetic wave propagation in randomly time-varying media,''
Nanophotonics \textbf{14}, 3317--3328 (2025).

%

\bibitem{tv9}
V.~Pacheco-Pe\~na and N.~Engheta, ``Temporal equivalent of the Brewster angle,''
Phys. Rev. B \textbf{104}, 214308 (2021).

%

\bibitem{sk1}
S.~Kim and K.~Kim, ``Propagation of Dirac waves through various temporal interfaces, slabs, and crystals,''
Phys. Rev. Res. \textbf{5}, 023162 (2023).

\bibitem{ok}
F.~Ok, A.~Bahrami, and C.~Caloz, ``Electron scattering at a potential temporal step discontinuity,''
Sci. Rep. \textbf{14}, 5559 (2024).

%

\bibitem{brewster1}
J.~E.~Sipe, P.~Sheng, B.~S.~White, and M.~H.~Cohen,
``Brewster anomalies: a polarization-induced delocalization effect,''
Phys. Rev. Lett. \textbf{60}, 108--111 (1988).

\bibitem{jord} T. M. Jordan, J. C. Partridge, and N. W. Roberts, ``Suppression of Brewster delocalization anomalies in an alternating
isotropic-birefringent
random layered medium,'' Phys. Rev. B \textbf{88}, 041105(R) (2013).

%

\bibitem{brewster4}
K.~J.~Lee and K.~Kim,
``Universal shift of the Brewster angle and disorder-enhanced delocalization of p waves in stratified random media,''
Opt. Express \textbf{19}, 20817--20826 (2011).

\bibitem{kim5}
K.~Kim and S.~Kim,
``Anderson localization and Brewster anomaly of electromagnetic waves in randomly-stratified anisotropic media,''
Mater. Res. Express \textbf{6}, 085803 (2019).

%

\bibitem{fang} A. Fang, Z. Q. Zhang, S. G. Louie, and C. T. Chan, ``Anomalous Anderson localization behaviors in disordered pseudospin systems,'' Proc. Natl.
Acad. Sci. U.S.A. \textbf{114}, 4087--4092 (2017).

\bibitem{kk1} S. Kim and K. Kim, ``Anderson localization and delocalization of massless two-dimensional Dirac electrons in random
one-dimensional scalar and vector potentials,'' Phys. Rev. B \textbf{99}, 014205 (2019).

\bibitem{kk2} S. Kim and K. Kim, ``Anderson localization of two-dimensional massless pseudospin-1 Dirac particles
in a correlated random one-dimensional scalar potential,'' Phys. Rev. B \textbf{100}, 104201 (2019).

%

\bibitem{lin1} Z. Gong, R. Chen, Z. Wang, X. Xi, Y. Yang, B. Zhang, H. Chen, I. Kaminer, and X. Lin, ``Free-electron
resonance transition radiation via Brewster
randomness,'' Proc. Natl.
Acad. Sci. U.S.A. \textbf{122}, e2413336122 (2025).

\bibitem{lin2} Z. Gong, R. Chen, H. Chen, and X. Lin, ``Anomalous Maxwell-Garnett theory for photonic time crystals,'' Appl. Phys. Rev. {\bf 12}, 031414 (2025).

%

\bibitem{bell}
R.~Bellman and G.~M.~Wing,
\textit{An Introduction to Invariant Imbedding}
(Wiley, New York, 1976).

\bibitem{gol}
M.~A.~Golberg, ``A generalized invariant imbedding equation,''
J. Math. Anal. Appl. \textbf{33}, 518--528 (1971).

\bibitem{ramm}
R.~Rammal and B.~Doucot,
``Invariant imbedding approach to localization. I. General framework and basic equations,''
J. Phys. (Paris) \textbf{48}, 509--526 (1987).

\bibitem{klya}
V.~I.~Klyatskin, ``The imbedding method in statistical boundary-value wave problems,''
Prog. Opt. \textbf{33}, 1--127 (1994).

\bibitem{kk0}
K.~Kim, D.~K.~Phung, F.~Rotermund, and H.~Lim,
``Propagation of electromagnetic waves in stratified media with nonlinearity in both dielectric and magnetic responses,''
Opt. Express \textbf{16}, 1150--1164 (2008).

\bibitem{sk2}
S.~Kim and K.~Kim,
``Invariant imbedding theory of wave propagation in arbitrarily inhomogeneous stratified biisotropic media,''
J. Opt. \textbf{18}, 065605 (2016).

\bibitem{sk3}
S.~Kim and K.~Kim,
``Giant overreflection of magnetohydrodynamic waves from inhomogeneous plasmas with nonuniform shear flows,''
Phys. Fluids \textbf{34}, 127108 (2022).

%

\bibitem{fur}
K.~Furutsu,
``On the statistical theory of electromagnetic waves in a fluctuating medium (I),''
J. Res. Natl. Bur. Stand. D \textbf{67}, 303--323 (1963).

\bibitem{nov}
E.~A.~Novikov,
``Functionals and the random-force method in turbulence theory,''
Sov. Phys. JETP \textbf{20}, 1290--1294 (1965).

%

\bibitem{mendon}
J.~T.~Mendon\c{c}a,
``Temporal Klein model for particle-pair creation,''
Symmetry \textbf{13}, 1361 (2021).

\bibitem{neto}
A.~H.~Castro~Neto, F.~Guinea, N.~M.~R.~Peres, K.~S.~Novoselov, and A.~K.~Geim,
``The electronic properties of graphene,''
Rev. Mod. Phys. \textbf{81}, 109--162 (2009).

\bibitem{peres}
N.~M.~R.~Peres,
``The transport properties of graphene: an introduction,''
Rev. Mod. Phys. \textbf{82}, 2673--2700 (2010).

\bibitem{ploc}
P. Plochocka, C. Faugeras, M. Orlita, M. L. Sadowski, G. Martinez, M. Potemski,
M. O. Goerbig, J.-N. Fuchs, C. Berger, and W. A. de Heer,
``High-energy limit of massless Dirac fermions in multilayer graphene using magneto-optical transmission spectroscopy,''
Phys. Rev. Lett. \textbf{100}, 087401 (2008).

\bibitem{lher}
A.~Lherbier,
``Transport properties of graphene containing structural defects,''
Phys. Rev. B \textbf{86}, 075402 (2012).

%

\bibitem{rech} M. C. Rechtsman, J. M. Zeuner, Y. Plotnik, Y. Lumer, D. Podolsky, F. Dreisow, S. Nolte, M. Segev, and A. Szameit,
``Photonic Floquet topological insulators,'' Nature {\bf 496}, 196--200 (2013).

\bibitem{plot} Y. Plotnik, M. C. Rechtsman, D. Song, M. Heinrich, J. M. Zeuner, S. Nolte, Y. Lumer, N. Malkova, J. Xu, A. Szameit, Z. Chen, and M. Segev, ``Observation of unconventional edge states in `photonic graphene','' Nat. Mater. {\bf 13}, 57--62 (2014).

\bibitem{fan} K. Fang, Z. Yu, and S. Fan, ``Realizing effective magnetic field for photons by
controlling the phase of dynamic modulation,'' Nat. Photonics {\bf 6}, 782--787 (2012).

\bibitem{cwang} C. Wang, M. Zhang, X. Chen, M. Bertrand, A. Shams-Ansari, S. Chandrasekhar, P. Winzer, and M. Lon\v{c}ar, ``Integrated lithium niobate electro-optic modulators operating at CMOS-compatible voltages,'' Nature {\bf 562}, 101--104 (2018).

%
































\end{thebibliography}
\end{document}